\documentstyle[12pt]{article}

\begin{document}

\newcommand{\APSREF}{1}
\newcommand{\LETTER}{0}
%
% macros for equation numbering and references
%
\def\daga#1{{#1\mkern -9.0mu /}}
\newcommand{\Eq}[1]{Eq.~(\ref{#1})}
\newcommand{\Ref}[1]{Ref.~\cite{#1}}
\newcommand{\beq}[1]{\begin{equation}\label{#1}}
\newcommand{\eeq}{\end{equation}}
\newcommand{\bdm}{\begin{displaymath}}
\newcommand{\edm}{\end{displaymath}}
\newcommand{\beqa}[1]{\begin{eqnarray}\label{#1}}
\newcommand{\eeqa}{\end{eqnarray}}
\newcommand{\bdma}{\begin{eqnarray*}}
\newcommand{\edma}{\end{eqnarray*}}

\ifnum\APSREF=1
\newcommand{\prd}[3]{Phys. Rev. D{\bf #1}, #2 (#3)}
\newcommand{\physrep}[3]{Phys. Rep. {\bf #1}, #2 (#3)}
\newcommand{\plb}[3]{Phys. Lett. {\bf B#1}, #2 (#3)}
\newcommand{\npb}[3]{Nucl. Phys. {\bf B#1}, #2 (#3)}
\newcommand{\np}[3]{Nucl. Phys. {\bf #1}, #2 (#3)}
\newcommand{\prl}[3]{Phys. Rev. Lett. {\bf #1}, #2 (#3)}
\newcommand{\rmp}[3]{Rev. Mod. Phys. {\bf #1}, #2 (#3)}
\newcommand{\ibid}[3]{{\em ibid.} {\bf #1}, #2 (#3)}
\newcommand{\astropj}[3]{Ap. J. {\bf #1}, #2 (#3)}
\newcommand{\annphys}[3]{Ann. Phys. {\bf #1}, #2 (#3)}
\newcommand{\repprp}[3]{Rep. Prog. Phys. {\bf #1}, #2 (#3)}
\newcommand{\jmp}[3]{J. Math. Phys.  {\bf #1}, #2 (#3)}
\newcommand{\zphysc}[3]{Z. Phys. C  {\bf #1}, #2 (#3)}
\newcommand{\ijmpa}[3]{Int. J. Mod. Phys. A  {\bf #1}, #2 (#3)}
\newcommand{\mpla}[3]{Mod. Phys. Lett. A{\bf #1}, #2 (#3)}
\newcommand{\jmpa}[3]{J.  Phys A: Math. Gen.  {\bf #1}, #2 (#3)}

\else
\newcommand{\prd}[3]{Phys. Rev. D{\bf #1} (#3) #2 }
\newcommand{\physrep}[3]{Phys. Rep. {\bf #1} (#3) #2 }
\newcommand{\plb}[3]{Phys. Lett. {\bf B#1} (#3) #2 }
\newcommand{\npb}[3]{Nucl. Phys. {\bf B#1} (#3) #2 }
\newcommand{\np}[3]{Nucl. Phys. {\bf #1} (#3) #2 }
\newcommand{\prl}[3]{Phys. Rev. Lett. {\bf #1} (#3) #2 }
\newcommand{\rmp}[3]{Rev. Mod. Phys. {\bf #1} (#3) #2 }
\newcommand{\ibid}[3]{{\em ibid.} {\bf #1} (#3) #2 }
\newcommand{\astropj}[3]{Ap. J. {\bf #1} (#3) #2 }
\newcommand{\annphys}[3]{Ann. Phys. {\bf #1} (#3) #2 }
\newcommand{\repprp}[3]{Rep. Prog. Phys.  {\bf #1} (#3) #2 }
\newcommand{\jmp}[3]{J. Math. Phys.   {\bf #1} (#3) #2 }
\newcommand{\zphysc}[3]{Z. Phys. C   {\bf #1} (#3) #2 }
\newcommand{\ijmpa}[3]{Int. J. Mod. Phys. A  {\bf #1} (#3) #2 }
\newcommand{\mpla}[3]{Mod. Phys. Lett. A{\bf #1} (#3) #2 }
\newcommand{\jmpa}[3]{J.  Phys A: Math. Gen.   {\bf #1} (#3) #2 }
\fi

\ifnum\LETTER=0
	\renewcommand{\theequation}{\thesection.\arabic{equation}}
	\newcommand{\newsection}[1]{\section{#1}\setcounter{equation}{
0}}
\fi

\newcommand{\aslash}[1]{{\rlap/#1}}
\newcommand{\splash}[1]{{#1\mkern -9.0mu /}}
% usage: \splash k
\newcommand{\splashh}[2]{{#2\mkern -#1mu /}}
% usage: \splashh{9}k

\baselineskip=20pt

%
% title
%

\begin{flushright}
Preprint IFUNAM:\\
FT95-70  .\\
hep-th/9501041
\end{flushright}

\vskip1.5cm

\begin{center}{\LARGE\bf
  Semilocal nontopological vortices in a Chern-Simons theory \\
\vskip0.2in
}\end{center}
\begin{center}
by

{\bf{ Manuel Torres\footnote{e-mail:
manuel@teorica1.ifisicacu.unam.mx  }
}}\\\vskip0.3in
{\small\it Instituto de Fisica,  UNAM,
Apdo. Postal 20-364, \\
\vspace{-2mm}
01000  Mexico, D.F., Mexico.}

\vskip1.0in
\noindent
{\bf{Abstract}}
\end{center}
\vskip0.05in
\setlength{\baselineskip}{0.2in}
We  show the existence of  self-dual semilocal  nontopological
vortices in
a  $\Phi^2$     Chern-Simons (C-S) theory.   The  model  of scalar
and
gauge fields
with  a  $SU(2)_{global}  \times U(1)_{local}$  symmetry  includes
both the C-S term and an anomalous magnetic contribution.
It is demonstrated here, that the vortices are stable or unstable
 according to whether
the vector topological mass $\kappa$ is less than  or
greater than the  mass $m$ of the scalar field.
At  the boundary, $\kappa = m$, there is a two-parameter family
of solutions  all saturating the  self-dual  limit. The vortex
solutions
continuously  interpolates between a ring shaped  structure and a
flux
tube  configuration.

\vskip1cm
\begin{center}
\noindent
\rule[.1in]{3.0in}{0.002in}
\end{center}
\vskip1cm
%PACS. numbers: 11.15.-$q$, 11.17.+$y$, 74.65.+$n$

\newpage
\baselineskip=20pt

%
% Section 1
%
\newsection{Introduction}\label{introduction}

Topological defects are  known to exist in theories with
spontaneously broken
symmetries.
For example, Nielsen and Olesen discovered that   Abrikosov type
vortices
(A-N-O)  appear
as classical solutions of an abelian Higgs model  \cite{nielsen}.
 These vortices carry magnetic flux but are
electrically neutral.
 Furthermore, for a $\Phi^4$  Higgs potential, and
when the parameters are chosen to make the
vector and scalar masses equal, minimum energy vortex configurations
arise
that satisfy first order differential equations  \cite{bogo,devega}.
In this
limit,
known as the Bogomol'nyi limit, the vortices become non-interacting
\cite{jacobs}.

In recent years the charged vortex solution  \cite{paul}  of the
Abelian Higgs
model in $(2 +1 )$ dimensions
 with  a Chern-Simons term  has attracted a lot of attention in the
literature,
because they can be considered as candidates for anyonlike objects
\cite{revjac}.
 These vortices were shown to exist even in the absence
of gauge field kinetic term (Maxwell term)  \cite{jatkar}. This
theory,  where
the kinetic action for the gauge field is solely the Chern-Simons
term is known as the pure C-S theory  (P-C-S)
\cite{deser,jackiwplano}. It was
recently shown
that  the P-C-S theory with an special choice of the scalar
potential $V(\Phi)$  supports self-dual  topological and
nontopological vortices  \cite{hong,jackwein}. For this theory,
$V(\Phi)$ is a
sixth-order potential, and the corresponding
Bogomol'nyi  limit is obtained when the vector and scalar masses
are equal.

More recently an Abelian Chern-Simons model which includes both the
\
C-S term and an anomalous magnetic contribution, in addition to the
Maxwell term, has been studied  \cite{torres}.
It was shown that for a special relation between the   C-S mass and
the anomalous magnetic coupling, the equations for the gauge fields
reduce from second- to first-order differential equations,
similar to those of the pure C-S theory. Furthermore,
it was demonstrated that nontopological  charged
vortices satisfy a set of
Bogomol'nyi-type equations for a quadratic potential
$V(\Phi) = (m^2/2) \Phi^2$, when $m$ and the topological masses
are equal. This model possess a local $U(1)$ symmetry, so we will
refer to it
as the
local $\Phi^2$ model.

On  the other hand, new interest in the study of string-like  defects
has
arisen after the observation made by Vachasparti and Ach\'{u}carro
\cite{vachasparti}  that  the Nielsen-Olesen  vortex solution  can be
embedded  into  a larger theory  which has  a global  $SU(N)$
symmetry in
addition  to the local $U(1)$
symmetry, these objects are  known  as semilocal  vortices. Semilocal
vortices
also appear
in Chern-Simons  theories; Khare \cite{khare}  obtained  semilocal
vortex
solutions
when  the local   $\Phi^6$ Abelian Higgs model  of   Refs.
\cite{hong,jackwein}
is extended to a semilocal one.

 In this paper
we consider a  $SU(2)_g \otimes U(1)_l$ model,  where only the
overall $U(1)$
phase is gauged
and  $SU(2)$ is a  global  symmetry.
 We will consider a
simple  $\Phi^2$ scalar potential, so we are necessarily in the
symmetric phase
of the theory;
nevertheless we  shall find  static minimum energy nonotopological
vortex configurations.  We will be interested in the stability  of
these
semilocal  vortices.
We will be able to  write a   lower bound for the energy (Bogomol'nyi
bound),
when the
scalar $(m)$  and the topological $(\kappa)$  masses are equal. The
lower bound is saturated  when the fields satisfy a set of  first
order
differential equations
(self-dual  or Bogomol'nyi equations).  The self-dual vortices are
neutrally
stable, but
it will be shown that the vortex solutions become stable or unstable
according
to whether $\kappa$  is less  than or greater than   $m$.
The semilocal $\Phi^2$ model exhibits a   richer spectrum of
solutions as
compared
with the local model. In particular the vortices can  assume both a
ring
shaped
structure (typical of the local    P-C-S  models)  as well as  a flux
tube
structure ( typical of the A-N-O vortices).   We should  also point
out that
the stability condition  $m > \kappa$ is  exactly
opposite to the one  obtained for the semilocal Nielsen-Olesen
vortices
\cite{hindmarsh}.

 The paper is organized as follows. In  Sec. 2 we present the
semilocal model
and discuss their
general properties.  In  Sec. 3 we derive the self-duality
(Bogomol'nyi)
equations.
 The derivation of  this self-duality  equations  is   much more
involved than
 the  original Bogomol'nyi derivation for the Nielsen-Olesen  model,
therefore we present it  with some detail.  Sec. 4 is devoted to the
study of
the cylindrically symmetric  vortex solutions.  First,  it is shown
that the
local $\Phi^2$ vortex solution can be embedded into the larger
semilocal
string;  then a generalization of the ansatz is presented,
this generalized solution  represents a  two-parameter family of
solutions all
saturating the
self-dual limit.    Section 5 deals with stability analysis of the
solutions.
Sec. 6  presents  a numerical  study of the self-dual solitons.
Concluding remarks comprise the final section.

%
% Section 2
%
\newsection{Model}\label{model}

The model consists of a complex doublet

\begin{equation}\label{phivec}
\Phi = \left (\matrix{
\phi_1\cr
\phi_2\cr
}\right )
\,  , \end{equation}
with only the overall phase gauged. The semilocal Lagrangian is

\begin{equation}\label{lagrangian}
  {\cal L} =  -{1 \over 4} F_{\mu \nu} F^{\mu \nu} \, +
   {\kappa \over 4} \epsilon^{\mu \nu \alpha} A_\mu F_{\nu \alpha} \,
   + {1 \over 2} |D_\mu \Phi|^2 -  \frac{m^2}{2} |\Phi |^2
  \,  ,
\end{equation}
where  $F_{\mu \nu} = \partial_\mu A_\nu - \partial_\nu A_\mu$,
 we use natural units $\hbar = c = 1$ and the Minkowski-space metric
is
$g_{\mu \nu}  = diag(+1,-1,-1); \, \mu  =(0,1,2). \,$
The covariant derivative  $D_\mu$ includes both the usual minimal
coupling plus the anomalous magnetic contribution:

\begin{equation}\label{dercov}
 D_\mu \Phi = \big( \partial_\mu - i e A_\mu - i {g \over 4}
    \epsilon_{\mu \nu \alpha} F^{\nu \alpha} \big) \Phi \, ,
\end{equation}
with $g$ the anomalous  magnetic moment \cite{paul2}.  Notice that
the
Lagrangian \Eq{lagrangian}  has a
global $SU(2)$ symmetry and a local $U(1)$ symmetry.   We should also
remark
that   it is an specific feature of a ($ 2 + 1$) dimensional world,
that a Pauli-type coupling  ($i.e.$, a magnetic coupling) can be
incorporated into  the covariant derivative,  even for spinless
particles
\cite{stern,torres}.
In fact, it was demonstrated in Ref.  \cite{kogan},  that radiative
corrections
can induce a magnetic coupling for anyons, that is proportional to
the
fractional spin.

The equations of motion for the lagrangian in  \Eq{lagrangian} are

\begin{equation}\label{eqsmot1}
D_\mu D^\mu \Phi
        =  -  m^2 \Phi
 \,  ,
\end{equation}
\begin{equation}\label{eqsmot2}
 \epsilon_{\mu \nu \alpha}  \partial^\mu \big [ F^\alpha +
{ g \over 2 e} J^\alpha \big]
     =  J_\nu - \kappa F_\nu
      \,  .
\end{equation}
The last equation has been written in terms of the dual field,
$ F_\mu \equiv {1 \over 2} \epsilon_{\mu \nu \alpha} F^{\nu \alpha}$,
and the conserved matter current $J_\mu = (\rho, \vec J)$ is given by

\begin{equation}\label{current}
 J_\mu = - {i e \over 2} \bigg [ \Phi^\ast D_\mu \Phi -
\Phi \big( D_\mu \Phi \big)^\ast \bigg]
   \,  .
\end{equation}

The energy momentum tensor is obtained by varying the curved-space
form
of the action with respect  to the metric

\begin{eqnarray}\label{tmunu}
   T_{\mu \nu}    & = &    \left( 1 - {g^2 \over 4} |\phi|^2 \right)
 \left( F_\mu F_\nu  -  {1 \over 2} g_{\mu \nu} F_\alpha F^\alpha
\right)
     \nonumber\\
 \nonumber\\
    & +  & {1 \over 2}\left( \nabla_\mu \Phi \left( \nabla_\nu \Phi
\right)^\ast
- g_{\mu \nu} \left[{1 \over 2} |\nabla_\lambda \Phi|^2 -  {m^2 \over
2}|\Phi|^2 \right]
+ H. c. \right)
   \,,   \end{eqnarray}
where $\nabla_\mu = \partial_\mu - i e A_\mu$ includes only the gauge
potential contribution.  Notice, that both the Chern-Simons  and
linear terms
in $g$
do not appear explicitly in $T_{\mu\nu}$. This is a consequence of
the fact
that  these
terms  do not make use of the space-time  metric tensor $g_{\mu\nu}$;
consequently,
when  $g_{\mu\nu}$ is varied  to produce $T_{\mu\nu}$  no
contributions  arise
from
these terms \cite{jackiwplano}.

For the selected scalar potential the  theory is always in a
symmetric phase.
In this
case the theory possess two propagating modes in the trivial sector
(excitations
around the vacuum): an scalar field excitation with mass $m$ and
a  massive vector  mode  with mass $\kappa$. Instead, if we would
consider
a broken symmetry  phase  the theory
would contain three propagating modes, because  the gauge field
acquire two
distinct masses due to the $P$ and $T$ violating terms
\cite{pisarski,boyanowski}.

%
% Section 3
%
\newsection{The self-duality equations}\label{selfdual}

There is a particular relation between the C-S mass and the anomalous
magnetic moment for which the \Eq{eqsmot2} for the gauge fields
reduce
from second- to first-order differential equations
\cite{stern,torres,latinsky},
similar to those of the P-C-S type \cite{jackiwplano}.
 To get this  limit  notice that if the  following relation holds

\begin{equation}\label{kappag}
 \kappa = - {2 e \over g}
   \,  ,
\end{equation}
then it is clear that the  \Eq{eqsmot2} are solved identically if we
choose
the first order ansatz

\begin{equation}\label{eqcs}
 F_\mu  = {1 \over \kappa} J_\mu
  \,    ,
\end{equation}
that  have the same structure as   the equations of the P-C-S theory
\cite{jackiwplano}.
We will refer to the previous conditions as the P-C-S limit. However,
we should
notice that
  the explicit expression
for $J_\mu$ differs from the usual expression of the P-C-S theory,
because
according
to \Eq{current} and \Eq{dercov}  $J_\mu$ receives contributions from
the
anomalous magnetic moment.
This  P-C-S
equations (\Eq{eqcs})  imply that any object carrying magnetic flux
($\Phi_B$)
must
also carry electric charge ($Q$), with the two quantities related
as $Q = - \kappa \Phi_B$. In what follows we shall work in the  limit
in which
\Eq{eqcs} and
\Eq{kappag} are  valid,
so we consider  \Eq{eqcs} as the equation of motion for the gauge
fields,
instead
of the \Eq{eqsmot2}.

In the  so called  Bogolmol'nyi limit    all the equations of motion
are known
to become first
order differential equations \cite{bogo};  furthermore,  it is
possible  to
write
the equations of motion as self-duality  equations. In reference
\cite{torres},
it
was shown that for  the    local  model,   it is possible to find
Bogomol'nyi-type
equations   for a quadratic potential $(m^2/2)\Phi^2$, when  $m$ and
the
topological
masses are equal.  In that case,  a rotationally invariant   vortex
ansatz  was
substituted in the
expression for the energy functional,  then the Bogomol'nyi equations
were
found
in terms of the reduced number of functions that appear in the
ansatz. However,
the self-duality
equations in terms of the original fields $\Phi$ and $A_\mu$  without
assuming
the rotational invariance  were no presented there.
In order to discuss  the vortex solutions in the extended semilocal
model we
require
to know the  form of the self-duality  equations.
The derivation of  this self-duality equations  is not as
straightforward as in the  original Bogomol'nyi derivation for the
Nielsen-Olesen  solutions,
therefore we present with some detail  the deduction of  the
self-duality
equations for the
 present model.

In order to obtain  the self-duality  equations of motion let us
consider  the
energy density ($T_{00}$) in the static limit  written  in terms of
the
magnetic
  $B={1\over 2}\epsilon_{ij} F^{ij}  $
and electric $E_i = F^{0i}$  fields, this yields:

\begin{equation}\label{t00}
T_{00} = {1 \over 2} \left( 1 - {e^2 \over \kappa^2} |\Phi|^2 \right)
\left(B^2
+  |\vec E|^2 \right)
 + {1 \over 2} \left [ |\nabla_0 \Phi|^2  + \left(\nabla_i
\Phi\right)^\ast
\nabla_i \Phi  \right]
+ {1\over2} m^2 |\Phi|^2
  \,  .
\end{equation}
We recall that the covariant  derivative
$\nabla_\mu = \partial_\mu - i e A_\mu$ includes only the gauge
potential contribution.
Now,  we notice that we can exploit  the $\mu = 0$  component of the
equation
\Eq{current}  together with the  C-S equations of motion \Eq{eqcs}
to
express
$A_0$ in terms of the magnetic field

\begin{equation}\label{a0}
A_0 = {\kappa \over e^2 |\Phi|^2} \left[ 1 - {e^2 \over \kappa^2}
|\Phi|^2
\right] B
  \,  .
\end{equation}
{}From this equation we can observe,  that  for a time-independent
vortex
solution, the  $A_0$ component
can not be set to zero  otherwise the magnetic flux would vanish.
Using the above equation,  the first and the  third terms  in
\Eq{t00} can be
added together
to obtain  the simplifying  result

\begin{equation}\label{aux1}
 {1 \over 2} \left( 1 - {e^2 \over \kappa^2} |\Phi|^2 \right) B^2 +
{1 \over 2}
|\nabla_0 \Phi|^2 = {1 \over 2} {\kappa^2 \over e^2 |\Phi|^2 }
\left( 1 - {e^2 \over \kappa^2} |\Phi|^2 \right) \, B^2
  \,  .
\end{equation}

Likewise, from the $\mu = i$ components of \Eq{eqcs} and
\Eq{current} we can
express
the   electric field  in terms of $\nabla_i \Phi$:

\begin{equation}\label{ei}
  E_i  = {1 \over \kappa \left(1 - { e^2 \over \kappa^2} |\Phi|^2
\right) }
\left(- {i e \over 2}\right) \epsilon_{ij}
 \left[ \Phi ^\ast  \nabla_j  \Phi - \Phi \left(\nabla_j
\Phi\right)^\ast
\right]
  \,  .
\end{equation}

In order to simplify the second  and fourth terms of the energy
density
\Eq{t00},   it  is
 convenient to define a new covariant derivative  $\tilde{D}_i$
according to
the relation

\begin{equation}\label{ndercov}
\tilde{D}_i =  \nabla_i + i H F_i = \partial_i - i e A_i + i H F_i
  \,  ,
\end{equation}
where we have  introduced  the auxiliary function $H$  defined as

\begin{equation}\label{h}
H = - { \kappa \over e |\Phi|^2}  \left[ 1 - {e^2 |\Phi|^2 \over
\kappa^2} -
\sqrt{1 - {e^2 |\Phi|^2 \over \kappa^2}} \right]
  \,  .
\end{equation}
Using the previous results (\Eq{ei} and \Eq{ndercov}),   we can now
find that
 the second and fourth terms in \Eq{t00} add together to give the
result

\begin{equation}\label{aux2}
  \left( 1 - {e^2 \over \kappa^2} |\Phi|^2 \right)  |\vec E|^2  +
\left( \nabla_i \Phi \right)^\ast \nabla_i \Phi
=\left( \tilde{D}_i \Phi \right)^\ast \tilde{D}_i \Phi
  \,  ,
\end{equation}
where  summation over latin indices is from $i= 1$ to $i=2$.

Employing the results of  \Eq{aux1} and \Eq{aux2} we can write down
the energy
$ E = \int T_{00} d^2x$ as

\begin{equation}\label{ener1}
E  = \int d^2x \left(  {\kappa^2 \over 2 e^2 |\Phi|^2   }
\left( 1 - {e^2 \over \kappa^2} |\Phi|^2 \right) \, B^2 +  {1\over2}
\left( \tilde{D}_i \Phi \right)^\ast \tilde{D}_i \Phi  + {1\over2}
m^2 |\Phi|^2
\right)
  \,  .
\end{equation}
The energy written in this form is similar to the expression that
appears in
the
Nielsen-Olesen model. Thus,   starting from \Eq{ener1} we  can
follow  the
usual
 Bogomol'nyi-type arguments in order to    obtain the self-dual
limit.  The
energy may then
be rewritten, after an integration by parts, as

\begin{eqnarray}\label{ener2}
   E      &= &   {1\over 2}   \int d^2x \left[
{ \kappa^2 \left( 1 - {e^2 \over \kappa^2} |\Phi|^2 \right) \over
e^2 |\Phi|^2
 } \left( B \mp
{e^2 |\Phi|^2  \over \left( 1 - {e^2 \over \kappa^2} |\Phi|^2 \right)
^{1/2} }
\right)^2
+  |\left(\tilde{D}_1 \mp i \tilde{D}_2 \right) \Phi|^2 \right]
           \nonumber\\
             \nonumber\\
            &+ & \int d^2 x  \left[ \left(m^2 - \kappa^2\right)
|\Phi|^2
\right]
\, \pm \, {\kappa^2\over e}  \oint_{r = \infty}{ d \vec l \cdot  \vec
\Omega}
  \,  \pm \,  { i \over 2} \oint_{r = \infty}{d \vec l \cdot \vec
\Lambda}
 \,   ,     \end{eqnarray}
where  the vectors $\vec \Omega$ and $\vec \Lambda$  are defined
according to
the relations:

\begin{eqnarray}\label{omega}
  \Omega_i    &= &  \left[ 1 - {e^2 |\Phi|^2 \over \kappa^2}
\right]^{1/2} A_i
   \, , \nonumber\\
\nonumber\\
 \Lambda_i   &= &   {\kappa^2 \over e^2 |\Phi|^2 }
\left( 1 -  \left[ 1 - {e^2 |\Phi|^2 \over \kappa^2}
\right]^{1/2}\right)\left(
\Phi^\ast \partial_i \Phi -
\Phi \partial_i \Phi^\ast \right)
   \,.    \end{eqnarray}

For any nontopological soliton the asymptotic conditions are such
that $\Phi
\to 0$ at spatial
infinity. Thus, the line integral of $\vec \Lambda$ in \Eq{ener2}
vanishes,
whereas the line
integral of $\vec \Omega$   yields     the magnetic flux

\begin{equation}\label{magf}
 \oint_{r = \infty}{ d \vec l \cdot  \vec  \Omega} \rightarrow
\oint_{r = \infty}{ d \vec l \cdot  \vec  A}   \, \equiv \Phi_B
  \,  .
\end{equation}
In what follows,  we consider only those configurations
that fulfill the condition  $|\Phi| \leq \kappa / e$.  Consequently,
from
\Eq{ener2} we
can conclude  that the energy is bounded below; for a fixed value of
the
magnetic flux, the lower  bound is given by   $E \, \geq{\kappa^2
\over e}
\Phi_B$ provided  that the potential is chosen
as a ${m^2 \over 2} |\Phi|^2$ with the critical value $m = \kappa$,
$i.e.$ when
the scalar and
the the topological masses are equal. Therefore,  in this limit  we
are
necessarily in the
symmetric phase of the theory. From \Eq{ener2} we  see that the lower
bound for
the energy

\begin{equation}\label{ener3}
 E \, = \, {\kappa^2 \over e} |\Phi_B| \, = \, {\kappa \over e} |Q|
  \,  ,
\end{equation}
is saturated when the following self-duality equations are satisfied:

\begin{equation}\label{selfdual1}
   \tilde{D}_1  \Phi  \,  =   \,  \pm i \tilde{D}_2 \Phi
  \,  ,
\end{equation}
\begin{equation}\label{selfdual2}
   B \,   =  \, \pm {e |\Phi|^2 \over \left[ 1 - {e^2 |\Phi|^2 \over
\kappa^2}
\right]^{1/2}}
  \,  ,
\end{equation}
where the  upper (lower) sign corresponds to positive (negative)
value of the
magnetic
flux.   We should remark that these self-duality are valid both for
the local
as well that
for the semilocal  $\Phi^2$ model.
The second of these equations implies  that the magnetic field vanish
whenever
$\Phi$
does.  For the local $\Phi^2$ model \cite{torres} the finiteness
energy
condition forces   the scalar field to vanish
both at the center of  the vortex and also at spatial infinity;
consequently
for  the local model  the
magnetic flux of the vortices  lies in a ring, rather than being
concentrated
at the
center as in the  A-N-O  solution.  As we shall see below,  for the
present
semilocal model
$\Phi$ does not vanish at the origin; furthermore, the vortices  can
assume
both a ring
shape  as well as a flux tube form.

With regard to
 the  self-duality  equations (\Eq{selfdual1} and  \Eq{selfdual2}),
we note that they are
   similar to those found
in  other  models, but  there are some important  differences.
First, we have  to point out  that   in order to derive the self-dual
limit it
was essential
to  introduce a   new covariant derivative $\tilde{D_\mu}$
(\Eq{ndercov})
that is
different   from the original    $D_\mu$ (\Eq{dercov}) that appears
in the
Lagrangian.
Of particular interest is to compare the present  results  with those
of the
Nielsen-Olesen  (A-N-O) and the pure Chern-Simons (P-C-S) models.
$(i)$ Similar self-duality equations  arise both in the local
\cite{bogo} and
semilocal \cite {vachasparti,hindmarsh}
A-N-O  models
 for a   potential of the form $(|\Phi|^2 - v^2)^2$  when the scalar
and vector
 masses are equal.  For the self-dual   A-N-O model   the
\Eq{selfdual} is
similar, but with
the normal covariant
derivative $D_\mu = \partial_\mu - ieA_\mu$ instead of
$\tilde{D_\mu}$;
furthermore,
\Eq{selfdual2} is replaced by one of the form $B \sim v^2 -
|\Phi|^2$, and  the
magnetic field
is maximum at the center of the vortex.
$(ii)$  Self-dual soliton solutions have also been found in  a  local
\cite{hong,jackwein}
   and
semilocal \cite{khare}  versions of the  P-C-S
  with no Maxwell term and no magnetic  moment contribution  for a
sixth order potential of the   form  $|\Phi|^2 (|\Phi|^2 - v^2)^2$,
when the
parameters
are chosen to make the vector and scalar masses  equal.
  Again the form of \Eq{selfdual1}
holds, but with the normal covariant derivative  $D_\mu =
\partial_\mu -
ieA_\mu$;
 whereas,  \Eq{selfdual2} is  now
replaced by one of the form  $B \sim |\Phi|^2 (v^2 - |\Phi|^2)$;
for this
kind of vortices
the magnetic flux is  localized within a ring around the center of
the vortex.

%
% Section 4
%
\newsection{Vortex solutions }\label{vortex}

The local  $\Phi^2$  vortex  solution of reference \cite{torres} can
be
embedded in one of the
components of  $\Phi$. Explicitly  for the semilocal  model a static
rotationally symmetric configuration of vorticity $n$ can be written
in
 in terms of  the  following ansatz:

\begin{eqnarray}\label{ansatz1}
  \vec A (\vec \rho) &= &  - \hat{\theta} { a(\rho) - n \over e\rho}
\, ,
\qquad A_0(\vec \rho) = {\kappa \over e} h(\rho) \, ,  \nonumber\\
 \nonumber\\
   \Phi(\vec \rho)  &=& {\kappa \over e} f(\rho) \exp{(- i n\theta)}
\left
(\matrix{
1\cr
   \cr
0\cr
}\right )
   \,,    \end{eqnarray}
where $(\rho, \theta)$ are the polar coordinates. If the previous
ansatz is
substituted into the
self-duality equations of motion   (\Eq{selfdual1} and
\Eq{selfdual2})
 the equations reduce exactly  to those of the  local
$\Phi^2$ vortices \cite{torres}, and hence we have  self-dual
nontopological
solitons
also in the present semilocal model. That this is the case,  may be
seen by
noting that if the lower component of  $\Phi$ is set  to zero the
resulting
model is exactly the local
$\Phi^2$ model. This means, that a solution of the local $\Phi^2$
model is
automatically a   solution of the semilocal model. Most of the
properties of
the semilocal  string   are
identical  to those of the local vortex: namely they represent
charged flux
tubes that carry
fractional spin as well as magnetization.  However, there may be a
difference:
the stability
of the semilocal vortex may be different from the stability  of the
local one.
Indeed, if one holds the lower component of $\Phi$  to zero and
perturbs only
the upper
component, it is exactly the same as perturbing the local $\Phi^2$
solution
that is
known to be stable   when $m > \kappa$ \cite{escalona}.
However,  we can also perturb the lower component
of $\Phi$, so we should check whether the semilocal vortex remains
stable for
$m > \kappa$.

Before proceeding to study   the stability of the semilocal vortices
we notice,
following
Hindmarsh \cite{hindmarsh}, that the ansatz \Eq{ansatz1} can be
generalized.
In fact,  the most general ansatz which maintains the cylindrical
symmetry  is

\begin{eqnarray}\label{ansatz2}
  \vec A (\vec \rho) &= &  - \hat{\theta} { a(\rho) - n \over e\rho}
\, ,
\qquad A_0(\vec \rho) = {\kappa \over e} h(\rho) \, ,  \nonumber\\
 \nonumber\\
   \Phi(\vec \rho)  &=& {\kappa \over e}\left (\matrix{
 f(\rho) \exp{(- i n\theta)} \cr
\cr
F(\rho)    \exp{(- i m\theta)}  \cr
}\right )
   \,.   \end{eqnarray}
The same as in reference \cite{hindmarsh} it is sufficient to examine
only  the
case $m=0$;
this  is equivalent to  add a cylindrically symmetrical perturbation
to the
lower component of the   ansatz \Eq{ansatz1}.
With this ansatz the self-duality equations of  motion \Eq{selfdual1}
and
\Eq{selfdual2} become

\begin{eqnarray}\label{selfdual3}
 { a^\prime \over r}   &= &  \mp { f^2 + F^2 \over  h}
\, ,
\nonumber\\
\nonumber\\
   f^\prime   &= &   \pm {a f  \over r h} \pm {n f F^2 \over r \left(
f ^2 +
F^2  \right) }
\left( 1 - {1 \over h}  \right)     \, , \nonumber\\
\nonumber\\
   F^\prime   &= &   \pm { \left( a - n\right) F  \over r h} \mp {n f
^2 F
\over r
\left( f ^2 + F^2  \right) }
\left( 1 - {1 \over h}  \right)
   \,,     \end{eqnarray}
where  we have  introduced the dimensionless variable $r=\kappa
\rho$,  primes
denote
differentiation with respect to $r$ and we  make use of \Eq{a0} and
\Eq{selfdual2}
 to solve for  the $\mu = 0$ component of the gauge field that yields

\begin{equation}\label{auxh}
h =  \left(1 - f^2 - F^2  \right)^{1\over 2}
  \,  .
\end{equation}
The boundary conditions are selected in such a way that the solution
\Eq{ansatz2}
 is non-singular at the
origin and gives  rise to a finite energy solution;  then, the
problem is to
solve \Eq{selfdual3}
subject to the following boundary conditions:

\begin{eqnarray}\label{boundary}
     a&= & n, \qquad f=0, \qquad F^\prime=0, \qquad at \qquad r=0
\, ,
\nonumber\\
      a&\to& -\alpha , \qquad f\to 0, \qquad F\to 0, \qquad as \qquad
r \to
\infty
   \,.    \end{eqnarray}
Notice that at spatial infinity the value $a(\infty) = - \alpha$ is
not
constrained. Consequently,
we shall see that
for nontopological solitons the magnetic flux  is not quantized, but
rather  it
is a continuous
parameter describing the solution.

The system of differential equations \Eq{selfdual3} looks rather
involved,
 however  the system can
be simplified  considerably  if we notice  that the equations for $f$
and $F$
are not
independent. In fact  it can be easily demonstrated, that subject to
the
boundary conditions
\Eq{boundary},  $f$ and $F$ are related according to the relation

\begin{equation}\label{relfF}
F = \pm \left({r_0 \over r}  \right)^{\pm n} f
  \,   ,
\end{equation}
where $r_0$ is an arbitrary parameter.  Exploiting this result, the
three
differential equations
(\Eq{selfdual3})   reduce to two; it is convenient to write them in
terms of
the functions
$a(r)$ and $h(r)$, this  brings the equations to the form

\begin{eqnarray}\label{selfdual4}
   a^\prime   &= &    r {\left( h^2 - 1 \right) \over  h}    \, ,
\nonumber\\
\nonumber\\
    h^\prime  &= &     {\left( h^2 - 1\right) \over  r  h^2}
 \left[ a - {n  \over 1  + \left({r \over r_0}\right)^{2 n} } \right]
   \,.    \end{eqnarray}
In what follows,  we  select the signs (upper signs in al the
previous
equations)
corresponding to positive magnetic flux $(n  >  0)$. The equations
for $n < 0$
are obtained  with the replacement $ a \to - a$, $f \to f$, $F \to F$
and
$h \to - h$.
This  previous system of equations is more amenable to be treated
numerically
and
various properties of the solutions can be discerned by general
considerations.
In particular the large  distance  behavior  of the solutions yields

\begin{eqnarray}\label{larger}
   a &= &  - \alpha + {C \over \left( \alpha - 1 \right)   r
^{2\alpha - 2}}
   + {\cal O}(r^{-4\alpha + 4}) \, , \nonumber\\
\nonumber\\
   h &=&  1 - {C \over r^{2\alpha} }  + {\cal O}(r^{-4\alpha})
   \,  ,   \end{eqnarray}
where $C$  is a constant. From these  asymptotic expressions we  see
that the
magnetic field,
$B =  a^\prime/r$, falls  off like $r^{-2\alpha}$. Likewise the
electric field
$| \vec E | \propto h^\prime$  falls off like $r^{-(2 \alpha +1) }$.
We  should
remark that although the theory  includes  gauge fields  with mass
$\kappa$,
the magnetic field departs from the usual
$e^{-\kappa r}$ asymptotic  behavior and becomes a power law.
 From the asymptotic
expression for $a(r)$ we notice that there should be lower bound on
the  values
of
$\alpha$. Indeed,  $\alpha > 1$ so the second term in the first of
\Eq{larger}
is subleading
compared with the first.
For  small $r$ a power-series solution gives

\begin{eqnarray}\label{smallr}
     a &= &  n + {\left(h_0^2 - 1 \right) \over 2 h_0}  \, r^2 +
{\cal O}(r^4)
\, , \nonumber\\
\nonumber\\
      h &= &  h_0  + {\left(h_0^2 - 1 \right) \over 2 h_0^2}
\left[ {\left(h_0^2 - 1  \right)\over 2 h_0} + {1 \over r_0^2}
\delta_{n,1}
\right] r^2  +{\cal O}(r^4)
   \,.    \end{eqnarray}
The constant $h_0$ is not determined by the behavior of the fields
near the
origin,
but  it  is instead  a parameter of the vortex solutions. The value
of $h_0$
may
be restricted by requiring  the proper behavior (\Eq{boundary}) as $r
\to
\infty$.

Once that the boundary conditions are known,  the topological numbers
of the
soliton
can be  explicitly computed.  With the ansatz  \Eq{ansatz2} the
magnetic field
is
$B= a^\prime/r$, and using the boundary conditions \Eq{boundary}  the
magnetic
flux yields

\begin{equation}\label{bflux}
\Phi_B = \int d^2x B = {2 \pi \over e }  \left( a(0) - a(\infty)
\right) =
 {2 \pi \over e }  \left( n + \alpha  \right)
  \,  ,
\end{equation}
notice that for nontopological solitons the  magnetic flux is not
quantized.
 The solutions are also characterized by the charge $Q$, spin $S$
(which is
general
fractional) and  magnetic moment $M$:

\begin{eqnarray}\label{topnumbers}
  Q &= &  - \kappa \Phi_B = {2 \pi \kappa \over e}  \left[ n + \alpha
\right],
\qquad S = {\pi \kappa \over e^2} \left( \alpha^2 - n^2  \right)
\, ,
\nonumber\\
\nonumber\\
  M &= &   - {\pi \over e}  \int_{0}^{\infty}{r^2 {dh \over dr} dr}
   \,.    \end{eqnarray}
Notice that the magnetic flux, the charge and the spin can be
calculated
explicitly,
because they  depend only  on the boundary conditions. Whereas,  the
magnetic
moment
 depends on the structure factor of the vortex configuration.

%
% Section 5
%
\newsection{Vortex stability }\label{vortexsta}

In the self-dual limit  the  soliton energy   is proportional to its
charge
(\Eq{ener3}).
This result is somehow  similar the  one obtained for the Q-balls
\cite{coleman};
although  in our model  the soliton  solutions are time-independent,
whereas
the Q-balls
are necesarily time-dependent.   For nontopological
solutions the soliton  charge is  of the same type
(Noether-charge) as that carried by the elementary excitations of the
theory .
Consequently,
we should  check wether the soliton is stable against  emission of
elementary
particles.
As we shall  see   at the self-dual point  ($m = \kappa$)  the
vortices are
neutrally stable.
For other values of the parameters the semilocal $\Phi^2$ vortices
are stable
when  $m > \kappa$ and are unstable otherwise.  Notice that this
situation  is
reversed  as
compared to the results obtained for the semilocal Nielsen-Olesen
(A-N-O)
vortices.
Indeed,  Hindmarsh \cite{hindmarsh}   demonstrated that the A-N-O
vortices are
stable when
the  mass of the scalar particle ($m$)  is smaller that the mass of
the  vector
 field ($m_v$).
We shall explain the reason of this difference at  the end of the
section.

As   stated above, the self-dual vortices are neutrally stable. This
fact can
be easily demonstrated
on account of  the  relation between the  energy and the charge
(\Eq{ener3}):
 $E =\kappa |Q|/e$. First, we recall that
the mass of the elementary excitations of the theory (scalar
particles)
is $m$ and the charge $e$.  Because of the charge conservation a
decaying soliton should radiate $Q/e$ ``quantas'' of the scalar
particles. Therefore, the energy of the elementary excitations
is ${\cal E} = m Q/e$. This indicates that the  vortices are at
the threshold of stability against decay to the elementary
excitations, due to the fact that  the ratio
$  {E_n / {\cal E}} = {\kappa / m}\, $
is  equal to one at the critical point $m = \kappa $.

In order to discuss the soliton stability we need to  discuss the
existence of
  two sums rules that can be derived from  \Eq{selfdual4}. To obtain
these sum
rules
we follow the reasoning introduced by Khare \cite{khare2}.
  The first sum rule
can be  established  by multiplying the first of \Eq{selfdual4}  by
$r$,
integrating over
$dr$ and using the asymptotic values of $a$, this yields

\begin{equation}\label{sumrule1}
 \alpha + n  =  \int_0^\infty   {r \left( 1 - h^2 \right) \over h} dr
  \,  .
\end{equation}
For the second  sum rule we multiply  the first  \Eq{selfdual4} by
$a$, after
integrating
and using  the second of \Eq{selfdual4}  we obtain

\begin{equation}\label{sumrule2}
 \alpha^2 -  n^2   = 2 n   \int_0^\infty  {1 \over 1 + \left({ r
\over r_0}
\right)^{2 n} }
{da \over dr} dr   + 2 \int_0^\infty  \left( 1 - h^2  \right) r dr
  \,  .
\end{equation}
{}From these two sum rules we an derive an upper limit  for
$\alpha$.   On account of \Eq{auxh} we   see that $0 < h < 1$; thus
we  can
combine
\Eq{sumrule1} and \Eq{sumrule2} to obtain

\begin{eqnarray}\label{sumrule3}
   \alpha + n   &= &   \int_0^\infty   {r \left( 1 - h^2 \right)
\over h} dr  >
\int_0^\infty   r \left( 1 - h^2 \right)  dr  \, , \nonumber\\
\nonumber\\
   &= &   {1 \over 2} \left( \alpha^2 - n^2  \right)
- n    \int_0^\infty  {1 \over 1 + \left({ r \over r_0}  \right)^{2
n} } {da
\over dr} dr
   \,.    \end{eqnarray}
{}From \Eq{selfdual4} we  notice that for the positive flux solution
$da/dr$ is
negative;
then,  the previous  equation yields  the  upper limit $\alpha < n +
2$.
As mentioned in the previous section, $\alpha$ is also bounded
below: $\alpha
> 1$.
Consequently, we find the bounds $1 <  \alpha < n + 2 $. This result
should be
compared
with the bounds  that were found for the local $\Phi^2$ model: $ n <
\alpha < n
+ 2$
 \cite{escalona}.
 In the limit $r_0  \to 0$    \Eq{sumrule2} implies the lower limit
$n <
\alpha$,
that is the same bound that was found  found for the  local  vortex
\cite{escalona}.

We now take up the discussion of the   the  vortex stability
for solution away from the self-dual point  ($m  \not= \kappa$).
We  shall follow   \cite{jatkar} and \cite{bazeia}   and  consider
the effect
of  perturbing the
potential around the  self-dual    point. We write the potential as

\begin{equation}\label{potmod}
 V (|\Phi| ) ={ (\kappa + \epsilon)^2 \over 2} | \Phi |^2
  \,   ,
\end{equation}
that corresponds to a scalar mass $m = \kappa + \epsilon$,
with $ |\epsilon| \ll 1$. The functions $h$, $f$, $F$ and $a$ will be
modified
from their  self-dual  configurations. We write them as:

\begin{eqnarray}\label{solmod}
  a(r)  &= &    a_{sd}(r) + \epsilon a_1(r) + {\cal O}(\epsilon^2)
\, ,
\nonumber\\
   f(r)  &= &     f_{sd}(r)  + \epsilon f_1(r)  + {\cal
O}(\epsilon^2)   \, ,
\nonumber\\
   F(r)  &= &   F_{sd}(r)  + \epsilon F_1(r)  + {\cal O}(\epsilon^2)
\, ,
\nonumber\\
   h(r) &= &    h_{sd}(r) + \epsilon h_1(r) + {\cal O}(\epsilon^2)
   \,.    \end{eqnarray}
{}From now on, we use the subindex $sd$ to denote the values of the
quantities in the self-dual  point. Therefore,  the functions
$a_{sd}(r)$ and
$h_{sd}(r)$ are solutions
of the differential equations \Eq{selfdual4} and   they satisfy the
sum rules
\Eq{sumrule1}
and \Eq{sumrule2}.
On substituting these expressions into  the ansatz  \Eq{ansatz2}  the
resulting
expressions for  the energy functional
\Eq{ener2},  to $O(\epsilon)$,    can be  recast in the form

\begin{equation}\label{ener4}
 E_n = {\kappa \over e } |Q| + \epsilon \,
{2 \pi  \kappa  \over e^2   }  \int_0^\infty r  \left( 1 - h_{sd}^2
\right) dr
  \,  .
\end{equation}
Notice that the charge (\Eq{topnumbers})  is modified as compared to
its
self-dual  value, because the asymptotic value $\alpha$ changes
$\alpha = \alpha_{sd} + \epsilon \alpha_1$, where $\alpha_1$ is
unknown. Nevertheless,  when one consider
the ratio of the soliton energy to the energy of the elementary
excitations $({\cal E} = m Q/e)$ we obtain that to order
$\epsilon$, $\alpha_1$ does  not contribute:

\begin{equation}\label{ener5}
 {E_n \over {\cal E}}  = 1 -  \epsilon {2 \pi \over  e Q_{sd} }
\left[  \left(n + \alpha_{sd} \right)  -  \int_0^\infty r \left(1 -
h_{sd}^2\right) dr \right]
  \,  .
\end{equation}
We notice that on account of  the sum  rule \Eq{sumrule1} the
quantity
inside the square brackets is positive (see \Eq{sumrule3}).
 Thus, we conclude that  when
$\epsilon > 0 \, ( m > \kappa)$  the soliton solution is stable
against
dissociation into free scalar particles. Whereas, the soliton is
unstable
for $\epsilon < 0 \, ( m < \kappa)$.

As stated before, the stability condition is  exactly opposite to
that obtained
for the
semilocal A-N-O vortices \cite{hindmarsh}.
The reason is traced  to the  fact  that,   although  the  semilocal
A-N-O
vortices are
not strictly    topological,
 they inherit  some topological properties of the local model; e. g
the boundary conditions are not modified if we vary the parameters of
the model
and
 the magnetic  flux is quantized.
Instead, in the $\Phi^2$ model both the local and semilocal vortices
are of
nontopological
nature.
 Using a  variational procedure  Hindmarsh \cite{hindmarsh} found
that the
A-N-O vortex
energy increases above the Bogomol'nyi limit when  $m > m_v$;  but
stability
requires
that the Bogomol'nyi limit must be saturated  a condition that
contradicts the
possibility
of having finite value for the magnetic flux..
 In the case of the semilocal $\Phi^2$ vortices
 \Eq{ener4} seems to suggest that the soliton energy also
increases above the Bogomol'nyi limit
 when $m > \kappa$;  but this is not necessarily true,
because   now the charge is not quantized and therefore  its value is
also
modified.
Furthermore,  for nontopological solitons
the relevant quantity  is the ratio   $ {E_n / {\cal E}}$ of the
vortex energy
 to the  energy of the  elementary excitations. For values of the
parameters
such that  $m > \kappa$
 the mass of the elementary excitation  increases     and
 the net  effect is that the ratio  $ {E_n / {\cal E}}$ decreases
(\Eq{ener5}), so it renders
the soliton stable.

%
% Section 6
%
\newsection{Numerical solutions}\label{numerical}
To complete the analysis  of  the previous sections  we present the
numerical
solutions of the self-duality  equations (\Eq{selfdual4})  with the
boundary
conditions
given  in \Eq{boundary}.  The solutions depend on two  parameters
$r_0$ and
$h_0$
(see \Eq{relfF} and \Eq{smallr}) that define the regular solutions.
Regular
solutions
exists in certain region of the parameter space ($h_0 \, vs \, r_0$);
 first  we recall that $ 0 < h_0 \leq 1$. Furthermore,  the values of
$r_0$ and
$h_0$
should be selected in such a way that the boundary conditions
\Eq{boundary} at
$r = \infty$ can be met.
In this regard our results may be summarized in Fig. 1, where we show
that
there are
three regions in parameter space.  In  region $(III)$ there  are no
acceptable
solutions
consistent with the boundary conditions at infinity (\Eq{larger}).
Both in
regions
$(I)$ and $(II)$ there  are acceptable vortex solutions, but the
properties of
these solutions
vary  from one region to the other.  In region $(I)$ the vortices
have a flux
tube structure with
the magnetic field  peaked at the origin;
whereas in region $(II)$ the vortices have a ring structure,  in this
case the
maximum of the magnetic
field is not at the center of the vortex. We explain these results
below, in
what follows
we consider the case of  isolated  vortex  $i.e.$  $n = 1$.

The origin  of the boundary between regions $(I)$ and $(II)$ can be
easily
understood
 from the expression for the   asymptotic behavior of the fields at
small $r$
\Eq{smallr}.
The field $h(r)$ starts  at the value $h_0$ at the origin and should
always
approach the
 value $ h \rightarrow \, 1$ at spatial  infinity. According to
\Eq{smallr} the
 slope
of   $h(r)$ at the origin can either  positive or negative.   Looking
at
\Eq{smallr} it is convenient
to define the critical value $h_c$ as
\begin{equation}\label{hcrit}
h_c = {\sqrt{1 + r_0^4} - 1 \over r_0^2}
  \,  .
\end{equation}
The boundary between  regions $(I)$ and $(II)$ corresponds to the
equation
$h_0 = h_c(r_0)$.  In region $(I)$  we have $h_0 < h_c$,  the slope
of  $h(r)$
at the
origin is positive and the function increases monotonically  from
$h_o$ at
$r = 0$   to
the value $h = 1$ at spatial infinity.  Instead, in region $(II)$
$h_0 >  h_c$
  the function
$h(r)$ decreases from $h_0$ until it reaches a minimum value
$h_{min}$,  and
from this point
it  increases   to the value $h(\infty) \rightarrow  1$. It is clear
that this
kind of solution
will be acceptable  if  $h_{min} >0$, otherwise the  equations become
singular,
see \Eq{selfdual4}. The boundary between regions $(II)$ and $(III)$
represents
the
points at which  $h_{min}  =0$; the values at this boundary  are
found
numerically.
In  Fig. 2 we plot $h(r)$ for a selected   $r_0 = 1$
and  several  values of the parameter $h_0$. For the selected value
of  $r_0 =
1$
and looking at   Fig. 1, we expect    to have acceptable solutions
for any
value
of  $h_0$; however,   the behavior of the solutions  should change as
$h_0$
cross the point
 $h_c  = 0.4142$.  For $h_0   < h_c$ the solution increases
monotonically,
whereas
for $h_0 > h_c$ the solution  develops  a minimum   away from the
origin.
These properties of the solutions are verified     in the   plots
presented in
Figs.  1 and 2.
 The corresponding  solutions for   the function $a(r)$ are
presented in Fig.
3. As expected from \Eq{selfdual4}  the function
$a(r)$  decreases monotonically from $a(0) = 1$ to the value $a = -
\alpha$ at
$r = \infty$.

In Fig. 4 we show profiles for the magnetic field $B = a^{\prime}/r$
as a
function of $r$
for   several values of $h_0$ and $r_0 = 1$.   It is useful to notice
that
using  \Eq{selfdual2},
\Eq{ansatz2} and \Eq{auxh}
the magnetic field can be written in terms of the field $h(r)$ as

\begin{equation}\label{bvsh}
 B(r) \, = \, {\kappa^2 \over e}  \,  { \left(1 - h^2(r) \right)
\over h(r)}
  \,  .
\end{equation}
Thus, as $h_0$ increases the  magnetic field at the origin $(B
\propto (1 -
h_0)/h_0 )$
decreases.  Furthermore, as explained above,  in region $(I)$  we
have  $h_0 <
h_c$
and the field  $h(r)$ increases
monotonically  with $r$; consequently the magnetic field
decreases monotonically  and the vortex has a flux tube structure.
However as
 the value of $h_0$  increases above  $h_c$ the magnetic  field
smoothly
diminishes at the
origin  and develops a maximum  away from the origin. Therefore in
region
$(II)$   the  magnetic field of  the vortex  is localized within a
ring around
the normal core.
We recall that from the  Chern-Simon equation (\Eq{eqcs})  the charge
distribution
is  proportional to  the magnetic  field.

In Fig. 5 we show the electric field  ($E  \propto h^\prime$)  for
the values
of the parameters
mentioned above.   We notice that in region $(I)$ the  electric field
has  a
maximum at finite
$r$ ( the minimum is reached at the boundaries),  however  for
parameters   in
region ($II)$ the field develops also a minimum  away from the
origin.

The vortex solutions are characterized by the quantum numbers that
were
calculated
in section  4, namely  magnetic flux (\Eq{bflux}), charge and spin
(\Eq{topnumbers}).
These quantities as well  as the soliton energy (\Eq{ener3}) depend
on the
asymptotic value
of  the field  $a(\infty) \rightarrow  - \alpha$. The value of
$\alpha$ is
in general a function of  $r_0$ and $h_0$.  We recall  from section
5 that
$\alpha$ is bounded $ 1 < \alpha < 3$
 ( for $n = 1$). Fig. 6 shows the behavior of $\alpha$ as a function
of $h_0$
for a fixed value of
$r_0 = 1$;  we find that  $\alpha$ approaches  the upper limit
$\alpha
\rightarrow   3$ as
$h_0 \rightarrow1$.  Instead,  in Fig. 7  we present the plot  of
$\alpha$  as
a function of
$r_0$  for $h_0 = 0.5$; in this case  as $r_0 \rightarrow \infty$ we
notice
that
$\alpha$ approaches the lower limit $\alpha \rightarrow 1$. For $r_0
< 0.69 $
there is no
vortex solution, in agreement with the results of Fig 1.

\newpage

%
% Section 7
%
\newsection{Conclusions}\label{conclusions}
By  extending the local $\Phi^2$ model of reference \cite{torres} to
a
$SU(2) \otimes U(1)$ semilocal model we found time independent
charged vortex
solutions. It was first necessary  to derive the self-duality
equations of
motion (\Eq{selfdual1})
in terms of the original scalar ($\Phi$) and vector fields ($A_\mu$).
This
self-dual limit
occurs for a simple $\Phi^2$ potential,  when the scalar and the
vector
topological
masses are equal.
 Although it is not possible to  find  exact  solutions  of the
self-duality
equations,  many of the elementary  properties of the vortices can be
uncovered.
The  self-dual  vortices are neutrally stable, but  including small
perturbations
away from  the self-dual point it was demonstrated that the
 vortices are stable or unstable
 according to whether
the vector topological mass $\kappa$ is less than  or
greater than the  mass $m$ of the scalar field.   It is interesting
to  remark
that this stability
condition is oposite to the one  obtained for the semilocal A-N-O
vortices
\cite{hindmarsh}.

The properties of these  vortices are indeed remarkable.
 Both  the local \cite{nielsen} and
semilocal \cite{vachasparti,hindmarsh}
 A-N-O  vortices   are known to have a flux tube structure with the
magnetic
field
peaked at the center of the vortex.  For  the  charged  vortices of
the   local
  P-C-S
 models    the magnetic field is concentrated  within  a ring
surrounding the
center of the vortex
with the magnetic field vanishing at this  point
\cite{hong,jackwein,boyanowski}.  Instead
in the semilocal   versions of the $\Phi^6$  P-C-S  model
\cite{khare}  the
magnetic field
is nonzero at the center of the vortex.
In the present  semilocal model the  vortex solution can have both a
ring
shaped structure
or a flux tube structure. The solution  continuously interpolates
between the
two configuration
as we move from region $(I)$ to region $(II)$ in parameter space
(Fig. 1).

There are several points to explore.
In particular the study  of the complete description of the
multisoliton
solutions and the
behavior away from the self dual limit deserve special attention.  Of
particular
interest  would be  to study the properties of Chern-Simons vortices
upon
quantization,
because they  can be considered as candidates for anyonlike objects
in
quasiplanar systems.

\newpage

{\bf FIGURE CAPTIONS}

{\bf FIG 1.} The crossover lines separating the three regions  in
parameter
space
           for the $n = 1$ vortex solution.  The boundary between
region $(I)$
and $(II)$ is
           obtained from  $h_0 = h_c(r_0)$ where $h_c$ is given in
\Eq{hcrit}.
In region $(III)$
    there are no acceptable soliton solution, because $h_{min} $
becomes
negative.

{\bf FIG 2.} A plot of  $h(r)$  as  a function of $r$ for a fixed
$r_0 = 1$
and values  of
                     the parameter $h_0$ of 0.25,0.40,0.50,0.75,0.95.
For a
vortex of vorticity
                     $n = 1$.

{\bf FIG 3.} A plot of  $a(r)$  as  a function of $r$ for a fixed
$r_0 = 1$
and values  of
                     the parameter $h_0$ of 0.40,0.75,0.95. For a
vortex of
vorticity
                     $n = 1$.

{\bf FIG 4.} The magnetic field in units of $\kappa^2/e$   for the $n
= 1$
vortex solution,
                      a fixed  $r_0 = 1$ and values  of
                     the parameter $h_0$ of 0.25,0.40,0.50,0.75,0.95.

{\bf FIG 5.} The electric  field in units of $\kappa/e$   for the $n
= 1$
vortex solution,
                      a fixed  $r_0 = 1$ and values  of
                     the parameter $h_0$ of 0.25,0.40,0.50,0.75,0.95.

{\bf FIG 6.}   Behavior of $\alpha$ as function of $h_0$ for $r_0 =
1$ and $n =
1$.

{\bf FIG 7.}   Behavior of $\alpha$ as function of $r_0$ for $h_0 =
0.5$ and $n
= 1$.

\end{document}